\documentstyle[epsfig,longtable]{aipproc}

\begin{document}

\title{YSO Jets and Molecular Outflows: Tracing the History of Star Formation}
 
\author{Adam Frank$^1$,}
\address{$^1$Department of Physics and Astronomy, Box 270171
University of Rochester, Rochester NY 14627-0171}

\maketitle
\def\ea{{\it et al.}~}

\begin{abstract} 

Collimated outflows from Young Stellar Objects (YSOs) can be seen as
tracers of the accretion powered systems which drive them.  In this
paper I review some theoretical and observational aspects of YSO
outflows through the prism of questions relating to the protostellar
source.  The issue I address is: can collimated outflows be used as
``fossils'' allowing the history of protostellar evolution to be
recovered?  Answering this question relies on accurately identifying
where theoretical tools and observational diagnostics converge to
provide unique solutions of the protostellar physics.  I discuss potential
links between outflow and source including the time and direction
variability of jets, the jet/molecular outflow connection, and the
the effect of magnetic fields.  I also discuss models of the
jet/outflow collimation mechanism.

\end{abstract}

\section*{Introduction}

Collimated outflows from Young Stellar Objects take the form of both
narrow high speed jets and bipolar molecular outflows
\cite{Boea97}.  The last two decades has witnessed rapid growth in our
understanding of these systems as hydrodynamic/hydromagnetic phenomena. 
Emission line diagnostics have allowed conditions in the jets to be
studied in considerable detail \cite{Hartiganea95}, \cite{Heathea96}
while numerical simulations have successfully cataloged the physics of jet propagation
\cite{Norman93}.  Recent HST images have revealed the morphology of HH
jets in such detail as to strongly constrain all future theoretical
models of YSO outflows (in what follows I will refer to collimated YSO
flows in general as ``outflows'' and will specificlly distinquish 
jets from molecular outflows when needed).

While the rapid progress of YSO collimated outflow studies is
heartening for jet enthusiasts it also holds the promise of providing
substantive answers to questions relating {\it directly} to the star
formation process itself. With an abundance of high resolution spectral
and imaging data the morphology, kinematics and microphysical state of
the outflows can be mapped out in detail. On the theoretical side
numerical models are now becoming sophisticated enough to include
accurate inventories of microphysical processes which produce
emission.  In this way fine grained comparison between outflow
observations and models is becoming possible. Strong constraints on
outflow models should allow researchers to look back to the source and
discriminate between competing scenarios of accretion processes
responsible for assembling the star.  Such a synergy between theory and
data will allow the details of accretion related processes to be
recovered from collimated outflows the way biological evolution is read
off the fossil record. The principal difficulty in this endeavor is to establish
which questions to ask.

Using collimated outflows as fossil tracers of protostellar physics can
succeed if researchers identify issues which provide a reasonably unique
and unambiguous bridge between the large scale outflows ($L \le 10^7
AU$) and accretion processes occuring on smaller scales ($L \le 10^2
~AU$, see Calvet,these proceedings).   Only those characteristics of the
jets imposed close to the star (or at the nozzle where
collimation occurs) will be useful.  It is important therefore to cleave
between features in the outflows imposed by accretion related
physics and those related to secondary mechanisms such as
interactions with the intercloud/interstellar medium. For instance,
characteristics which result from dynamical instabilities can only
provide insights into accretion process when they can be linked to
properties near the central object \cite{StoneHard97}, \cite{Rossiea97}.

In what follows I attempt to identify some issues currently under study
that might establish the needed bridge between
outflows and the accreting protostar.  This list is by no means
complete and I apologize in advance for neglecting, due to lack of
space, many important investigations contributing to what might be called
``YSO outflow paleontology''.  

\section*{Observations: HH Jet Behavior} 

Fig 1 (taken from Reipurth {\it et al.} 97) shows HST images of three
images of jets with HH objects:  HH111, HH34, and HH46/47.  These systems
illustrate key features common to many HH object/jets \cite{Bo97}.

All three jets show spatial variability on a variety of scales.  In
particular, closely spaced chains of knots are visible in all three
systems. Sequences of larger, more widely separated features showing
distinct bow shock morphologies, also occur in all three jets. If the
spatial variations result from time variations inherent to the jet
source then the knots and bowshock chains give timescales of $\Delta
t_{knot} \approx 1 - 10 ~y$ and $\Delta t_{bow} \approx 100 - 1000 ~y$
respectively \cite{Rayea96}, \cite{Heathea96}. Much longer chains
of HH objects have recently been discovered by Bally and collaborators
(these proceedings). These "super-jets" can extend out to distances 
on the order of a few parsecs.  
Dynamical ages for the super-jet systems can be as large as $t \approx 10^5 ~y$, comparable to the time it takes to assemble the protostellar sources.

While HH111 and HH30 are remarkably straight, HH47 shows evidence of
direction variability on different scales.  Note in particular how the
jet beam in HH47 appears to connect to the side of the bright bow shock.  Both
molecular outflows \cite{Guea95} and the super-jets show evidence for
direction variability in the form of either precession (which produces
point-symmetric {\bf S}-shaped configurations) or random variations in
the form of jet ``wandering''.

Proper motion studies show the gas in these systems is traveling at
$v_j \approx ~300 ~km/s$.  Spectral line diagnostics indicating the
level of excitation and ionization in the shocks yield speeds of
order $v_{s} < 100 ~km/s$ \cite{Hartiganea95}.  The disparity between these two speeds
strengthens the case for very long jet lifetimes as they indicate the
visible portions of the jet are propagating into invisible jet material
ejected earlier.

The velocity and direction variability identified above may yield the
strongest points of contact relating large scale outflows to processes
at the source.  In the next two sections I review theoretical models
which attempt to establish an explicit link between jet propagation
physics and variability at the source.

\begin{figure} 
\centerline{\epsfig{file=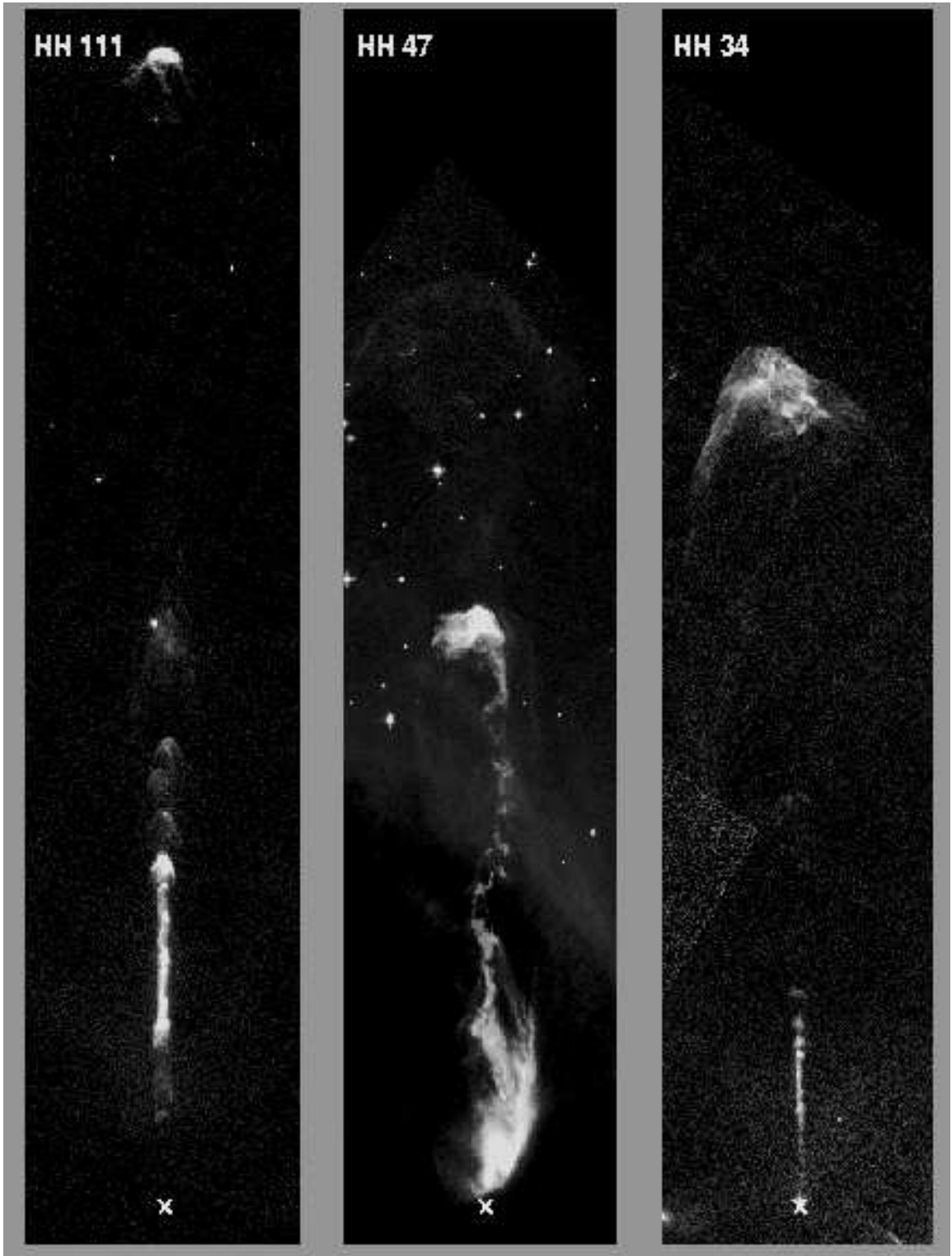,width=6in,height=8in}}
\vspace{0pt}
\caption{
Combined H$\alpha$ and [SII] emission maps of three HH jet systems
from Reipurth {\it et al.} 1997.
}\label{myfirstfigure}
\end{figure}

\section*{Jet Variability and The Molecular Outflow Connection}

Numerical simulations have taken the study of YSO jets into maturity. 
Jets can be decomposed into 4 main dynamical elements:  the {\it beam}
of unshocked jet material; the {\it jet shock} which decelerates
material in the jet beam; the {\it bow shock} which accelerates
ambient material; the {\it cocoon} decelerated jet gas surrounding the
beam;  The pioneering work of Blondin, Fryxell \& Konigl (1990)
demonstrated the dramatic effect that collisionally excited radiative losses
have on jet propagation dynamics. Their simulations showed that heavy
($\eta = \rho_j/\rho_a > 1$, $a$ = ambient, $j$ = jet) ``radiative
jets'' differ substantially from  ``light'' ($\eta < 1$) extra-galactic
jets. In radiative jets the loss of pressure support in post-shock
regions collapses the bow and jet shocks into a thin dense shell which
quickly fragments.  The pressure in the cocoon is also reduced. Thus,
unlike light jets, the cocoon in a radiative jet does not drive pressure
waves into the beam allowing it to remain relatively undisturbed.

Simulations which include velocity variability at the source have been
explored by a number of authors, (\cite{SN93}, \cite{DPB94},
\cite{BiRa94}) including a recent, noteworthy investigation by Suttner
\ea 1997.  These studies demonstrate that periodic velocity pulses
produce multiple ``internal working surfaces'' downstream in the beam
as faster material catches up to, and shocks slower moving gas. These
results argue that the widely separated multiple bow shocks observed in
YSO jets can be attributed directly to temporal variations at the jet
source.  The situation is less clear for the narrowly spaced chains of
knots.  Recent studies of Kelvin-Helmholtz instabilities in YSO jets
have shown that small scale periodic structures can be imposed in the
jets via dynamics inherent to the beam itself (\cite{StoneHard97}
\cite{Rubiniea97}).  On the
other hand simulations of magneto-centrifugal jet production
\cite{OP97} show short time-scale variations imposed in the beam due to
processes inherent to the collimation process.

The connection between jets and Molecular Outflows remains one of the
most important issues facing YSO studies \cite{Cabitea97}.   Like the
jets, bipolar molecular outflows are ubiquitous in protostellar environments
and many YSOs exhibit both phenomena.  After almost two
decades of study it is still unclear if the the jets and outflows are
causally related or simply co-extensive.  Competing models invoke
either a ``wide-angle wind'' \cite{Shuea91} or a jet 
\cite{MassCher94} to drive the molecular outflows.  Discriminating between these models has
direct impact on issues relating to the physics of the protostar.

While some molecular outflows are quite narrow, many show poor
collimation with rather wide lobes and low aspect ratios.  Among other
problems, jet driven models have difficulty explaining the variety of
outflow shapes. This is an issue where variations in jet direction may
have a direct bearing. The effect of variations in jet direction has
been studied in precessing or wandering jet models.  Raga Canto \& Biro
(1993) argued that changes in the jet direction would produce weak
"sideways shocks" propagating across the jet cross section which may be
observable.  Using 3-D simulations Cliffe, Frank \& Jones (1996) confirmed
this prediction and demonstrated that direction variability leads to
significant deceleration of the leading sections of the jet. By
including a calculation of microphysical processes Suttner \ea (1997) have
shown that jets with precession and velocity variations imposed at the
source produce kinematical patterns well matched with observations.

Both Cliffe, Frank \& Jones (1996) and Suttner \ea (1997) have taken some
steps in addressing the width of jet driven molecular outflow lobes.
Based on earlier suggestions \cite{MassCher94}, these authors demonstrated
that precessing jets will produce a wide global bow shock enveloping the
entire ``corkscrew'' of the jet. If this global bow shock structure can
be identified with molecular outflows it may allow jet driven models
to recover both narrow and wide outflow lobes as the result simply
depends on the precession or wandering angle.  Velocity variations
\cite{RaCa93} in the beam have also been used as a means for widening
jet driven outflows.  In these models the bow shock is inflated via
pressure from gas``squirted'' out the sides of internal shocks.

\section*{Magnetized Radiative Jets} 

Hydrodynamic jets with and without radiative losses have been
well-studied.  Magnetized jets subject to collisionally excited
radiative losses have not, as yet, received extensive scrutiny. Given
the consensus that MHD processes are responsible for producing
astrophysical jets this represents a rather large gap in our
understanding of YSO outflows systems and constrains our ability to make
connections with protostellar physics. If the collimation process at
the source is MHD dominated then magnetic fields will remain embedded
in the jets as they propagate.  In particular both disk-wind (\cite{OP97},
see references therein)
and X-wind (Ostriker, these proceedings) models of jet collimation
produce jets with strong toroidal ($B = B_\phi$) fields.  The strength of the
jet fields can be seen in the relevant Mach numbers.  While sonic Mach
numbers of MHD collimated jets may be high ($M_s > 10$) the fast mode
Mach numbers are quite low ($M_f \approx 3$, \cite{Camen97}).

Direct observation of magnetic fields in protostellar jets would help
clarify issues surrounding jet origins.  Unfortunately such
measurements have generally proven to be difficult to obtain
\cite{Rayea97}.  A promising alternative is look for less direct
tracers of strong fields in YSO jets.   If jets are produced via MHD
processes, dynamically significant magnetic stresses should affect the
beam and jet head as they interact with the environment.  Thus the
propagation characteristics of protostellar jets may hold important
clues to their origins.

Cerqueira {\it et al.} (1997) and Frank {\it et al.} (1997) have recently
reported the first multi-dimensional simulations of radiative MHD
jets.  Both studies demonstrate that propagation characteristics of
jets with strong toroidal fields  differ significantly
from their pure hydrodynamic cousins.  This is illustrated in Figure 2
(from Frank {\it et al.} 1997) which compares the results of a weak field
and a
strong field radiative jet simulation.  Comparison of the two cases
demonstrates the effect of strong toroidal fields and cooling. 
The weak field jet exhibits a dense
shell and a relatively undisturbed jet beam.  In the strong field case
however ``hoop'' stresses associated with the radially directed
magnetic tension force inhibit sideways motion of shocked jet gas.
Material that would have spilled into the cocoon is forced into the
region between the jet and bow shock, forming a ``nose-cone'' of
magnetically dominated low $\beta$ gas ($\beta = P_g/P_B = 8 \pi
P_g/B^2$) \cite{Norman93}. Hoop stresses also collapse the beam near the nozzle,
producing strong internal shocks due to magnetic pinches.

Fig 2. shows that the weak and strong field radiative jets look
dramatically different from each other in terms of the morphology of
the jet head and beam.  In the weak field case there are no shocks in
the beam.  The strong field case shows multiple shock reflections . The
head of the weak field jet is quite ``blunt'' compared with the
strongly tapered strong field jet. The kinematics of the two jets also differ.
The average propagation speed for
the head of the strong field jet is $\approx 40\%$ higher than the weak field
case, even though both simulations have the same jet/ambient density
ratio, $\eta$.  The combination of higher shock speeds and strong pinch
forces in the radiative strong field jet produces a higher compression ratio
in the head of strong field jet.
If these results are born out in more detailed studies, particularly
those in 3-D, then indirect diagnostics of the presence of dynamically
strong fields in jets should exist.

\section*{Jet Collimation}

Of course the best way to link outflows with their protostellar
sources is to understand exactly how outflows are created.
Obscuration and resolution issues make observations of the inner
collimation regions difficult but progress is being made.  A number of
studies have provided strong evidence that jet collimation must occur
on observational scales of order $R \le 10 AU$ (\cite{Rayea96}).

The current consensus in the astrophysical community holds that YSO
jets are launched and collimated by MHD processes.  The
most popular models rely on magneto-centrifugal forces in either an
accretion disk (``Disk-winds'') or at the disk-star boundary 
(``X-winds'').  These studies have been quite successful in articulating the
physical properties of MHD collimation processes.  Indeed numerical
simulations \cite{OP97} have recently demonstrated the
ability of disk-wind models to produce both steady and time dependent
jets (Fig 3).  Models which rely on the interaction of a dipole
stellar field tied to an accreation disk have also shown promise in
both analytic and numerical studies (\cite{Goodsonea97} 
and references therein).

In spite of this success many questions remain.  The existence of
variety of working models leads to a basic uncertainty as to which
mechanism YSOs choose if MHD is the dominant launching {\it and}
collimation process.  There are also questions as to the effectiveness
of {\it collimation} in MHD models.  As Shu \ea 1995 point out MHD
collimation can be a slow process occurring logarithmically with height
above the disk.  A different set of numerical simulations \cite{Romanova97}
found that while the magneto-centrifugal process was effective at
launching a wind, it did not produce strong collimation of the wind
into a jet. 

Along with these issues, recent numerical studies have shown that pure
hydrodynamic collimation can be surprisingly effective at producing
jets.  Frank \& Mellema (1996) and
Mellema \& Frank (1997) have demonstrated that isotropic or wide angle
YSO winds interacting with toroidal density environments produce
oblique inward facing wind-shocks.  These shocks can be effective at
redirecting the wind material into a jet (see Fig 3).  If the wind from
the central source is varying this ``shock focusing'' mechanism can, in
principle, collimate jets on the observed physical scales.
 Similar mechanisms have been shown to work to in other
jet-producing contexts, particularly Planetary Nebulae. 
\cite{Frea96} \cite{BBH97}.

The variety of models currently available to produce YSO jets begs the
question of which observational diagnostics are needed to distinguish
the processes actually producing the jets. Further progress will come
as the collimation models mature and are capable of prediciting
observational consequences (ie from $B$ fields or shocks) in observable
regions of the flow. Thus critical observations concerning the
formation of the jets will, once again, have to come from downstream of
the collimation regions, i.e.  from the jets themselves.

\section{Summary} 

The issues cited in this paper are associated with outflows. How
do these issues specifically relate to questions inherent to the
physics of accretion? The time variability of jets relates to the
time-dependence of accreation, the FU Ori outbursts being a notable
example.  The direction variability of jets relates to the global
dynamics and stability of accretion disks.  Livio \& Pringle 1997, for
example, have shown that radiation induced warping of disks may lead to
precession in magneto-centrifugal jets. The presence and structure of
magnetic forces in jets relates to the existence and form of large
scale fields in the disks. If nose-cones do not occur in real YSO jets
then perhaps mechanisms which rely on strong toroidal fields are
excluded.

Thus YSO jets and outflows offer a unique opportunity for the study of
accretion powered systems.  Protostellar outflows can be observed with
exquiste detail in a variety of wavelengths including diagnostic
spectral lines.  The quality of the data combined with the long
lookback time inherent to the outflows offers the possibilty that a
large fraction of individual protostar's history might be recovered if
we learn were and how to look.  We are a long way from this now but the
prospect of having such capabilities is very exciting.


\begin{figure} 
\centerline{\epsfig{file=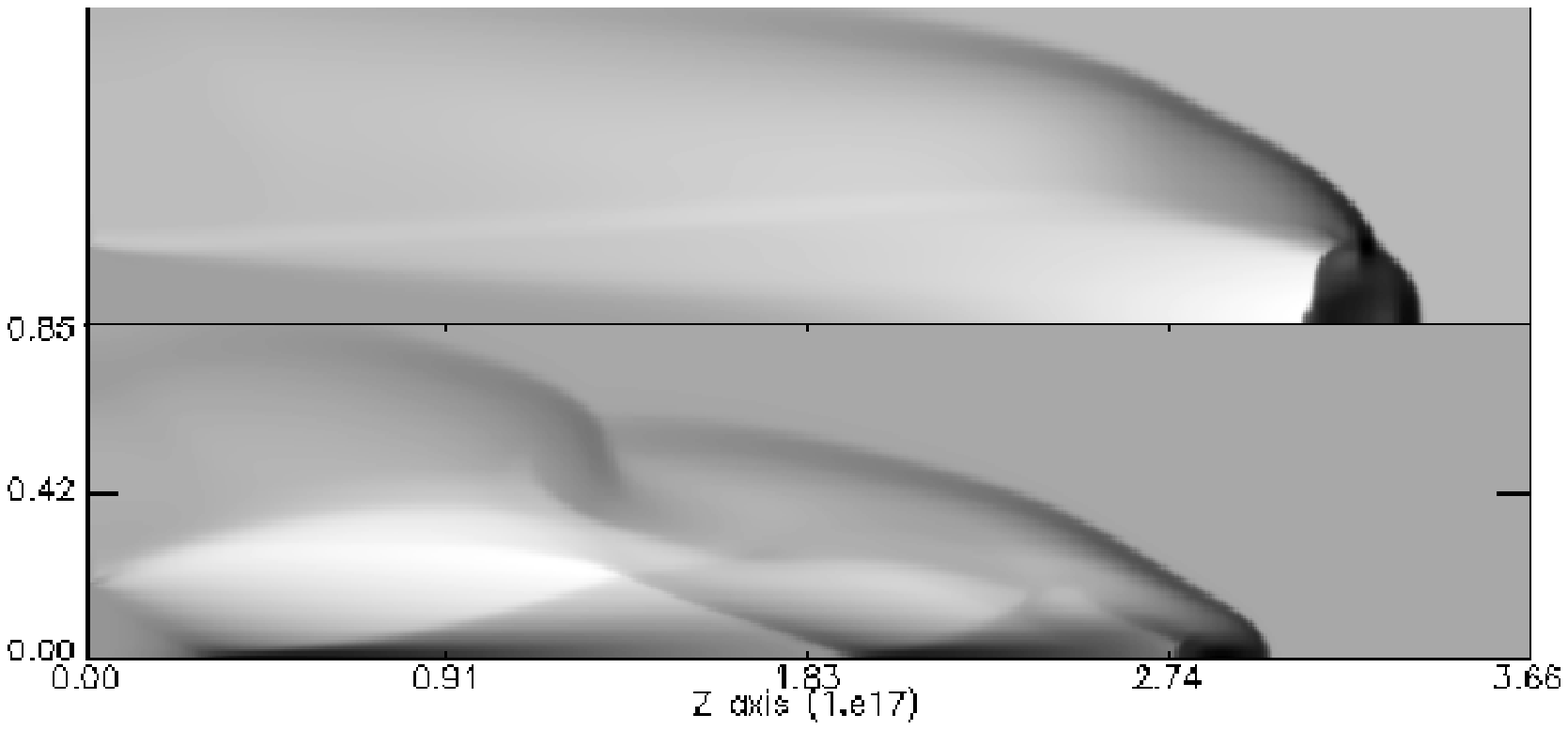,width=5in,height=5in}} 
\vspace{0pt}
\caption{Comparison of two YSO Jet simulations from Frank {\it et al}
1997.    Top: Grey-scale density map from a weak field radiative model 
(at t = 1,971
y).  Bottom: Grey-scale density map from a strong field radiative model 
(at t = 1,460 y). Both jets have
sonic mach numbers $\approx 10$. The strong field jet has a fast mode
mach number $\approx 5$. Note that narrow bow shock and strong internal
beam shocks in the strong field case.} 
\label{myfirstfigure}
\end{figure}

\begin{figure} 
\centerline{\epsfig{file=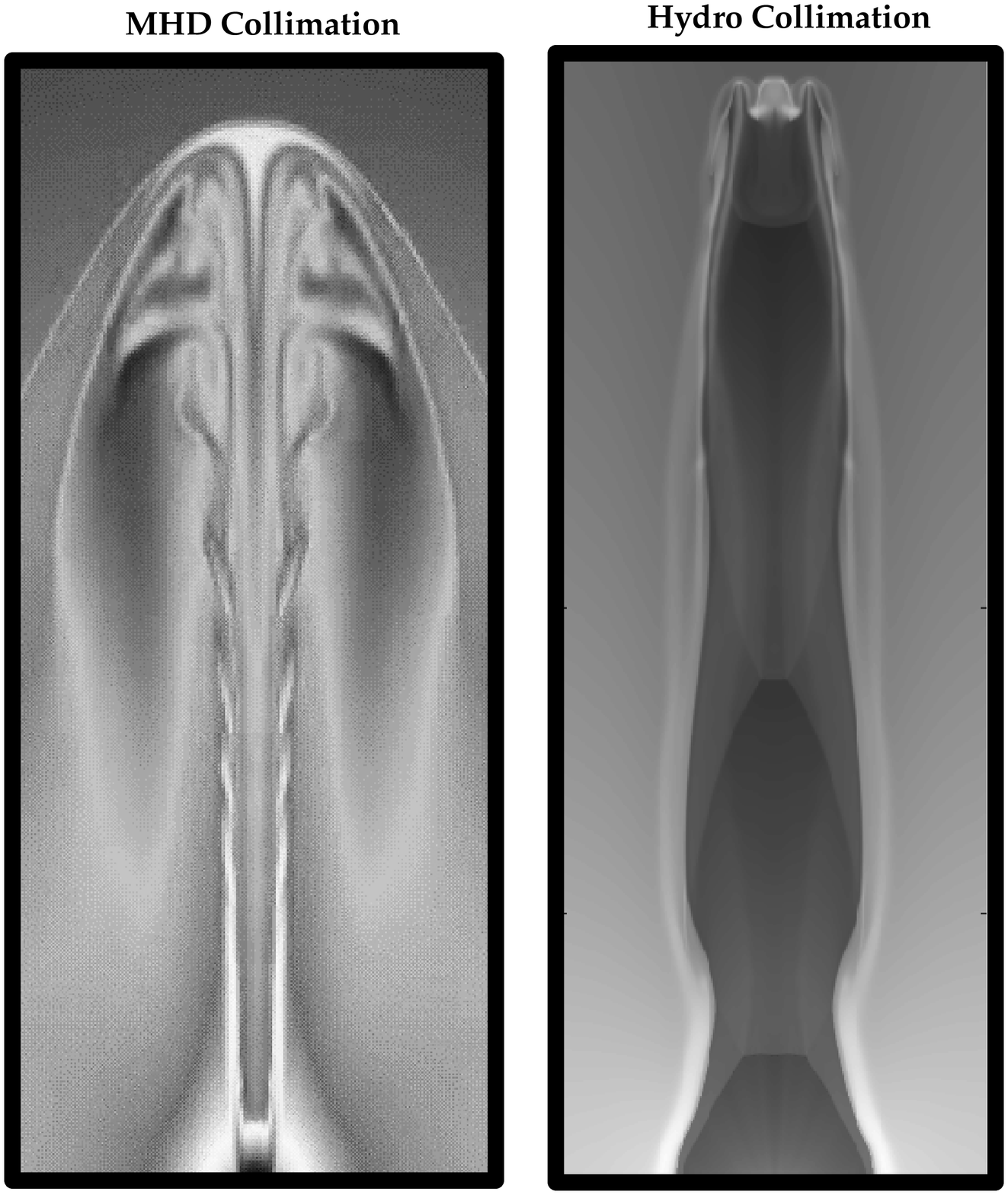,width=5in,height=6in}}
\vspace{0pt}
\caption{Comparison of MHD and Hydrodynamic Jet 
collimation models.  Left panel: greyscale representaion of density from
a MHD disk wind simulation (Ouyed and Pudritz 1997). Right:
greyscale representaion of density from
a hydrodynamic collimation model (Mellema \& Frank 1997).  
In the hydrodynamic model a spherical wind from the protostar interacts
with a toroidal density distribution producing a supersonic jet.}
\label{myfirstfigure}
\end{figure}

\end{document}